\documentclass[a4paper,11pt]{article}

\usepackage{pos}
\usepackage{graphicx}
\usepackage{svg}
\usepackage{caption}
\usepackage{subcaption}

\newcommand{\eV}{\; \text{eV}}
\newcommand{\MHz}{\; \text{MHz}}
\newcommand{\Xmax}{$X_{\text{max}}$}
\newcommand{\gcm}{\; \text{g}/\text{cm}^2}

\newcommand{\slice}[1]{#1_\text{slice}}

\title{Template synthesis approach for radio emission from extensive air showers}

\author*[a]{Mitja Desmet}
\author[a]{Stijn Buitink}
\author[b]{David Butler}
\author[a,b]{Tim Huege}
\author[b]{Ralph Engel}
\author[a,c]{Olaf Scholten}

\affiliation[a]{Inter-University Institute For High Energies (IIHE), Vrije Universiteit Brussel (VUB),\\
  	Pleinlaan 2, 1050 Brussels, Belgium}

\affiliation[b]{Institut für Astroteilchenphysik (IAP), Karlsruhe Institute of Technology (KIT),\\
	PO Box 3640, 76021 Karlsruhe, Germany}

\affiliation[c]{Center for Advanced Radiation Technology (KVI), University of Groningen,\\
	Zernikelaan 25, NL-9747 AA Groningen, The Netherlands}

\emailAdd{mitja.desmet@vub.be}
\emailAdd{stijn.buitink@vub.be}
\emailAdd{tim.huege@kit.edu}
\emailAdd{ralph.engel@kit.edu}
\emailAdd{scholten@kvi.nl}

\abstract{We present a novel way to synthesise the radio emission from extensive air showers. It is a hybrid approach which uses a single microscopic Monte-Carlo simulation to generate the radio emission from a shower with a different longitudinal evolution, primary particle type and energy. The method employs semi-analytical relations which only depend on the shower parameters to transform the radio signal in the simulated antennas. We apply this method to vertical air showers with energies ranging from $10^{17} \; \text{eV}$ to $10^{19} \; \text{eV}$ and compare the results with CoREAS using two different metrics. In order to gauge the performance over our simulation set, we subsequently use every shower in the set as a template to synthesise the emission from the other showers. Depending on the scoring metric, template synthesis reconstructs the radio emission with an accuracy of 5 to 10\%.}

\FullConference{%
  9th International Workshop on Acoustic and Radio EeV Neutrino Detection Activities - ARENA2022\\
  7-10 June 2022\\
  Santiago de Compostela, Spain}

\begin{document}
\maketitle

\section{Introduction}

The next generation of radio telescopes such as LOFAR 2.0 \cite{VanHaarlem2013} and SKA \cite{Dewdney2009} will collect more, as well as higher-quality data from extensive air showers (EAS). While this will allow to reconstruct the EAS in much greater detail, we pay a hefty price in computational cost. In order to reconstruct the shower parameters with radio we rely on dedicated sets of microscopic simulations, using the CORSIKA code with its CoREAS extension \cite{Heck1998, Huege2013}. This requires significant computation time, especially for large arrays as the calculation of the radio emission scales linearly with the number of antennas.

Therefore, to take full advantage of future detectors, we need a faster way to simulate the radio emission from EAS. To reach this goal, we follow the template synthesis procedure as outlined in \cite{Butler:2019xB}. In this contribution, we present an independent verification of the method as well as some enhancements. These include more robust fitting procedures and the implementation of scoring metrics to quantify the reconstruction quality.

Our approach, called template synthesis, consists of characterising the radio emission using semi-analytical expressions which are functions of the shower age. Employing these relations we can transform the radiation from a template shower to one with different parameters. The template is a single microscopic simulation, which takes care of all the geometrical effects and signal phases. From it, we can synthesise the emission from showers with an arbitrary longitudinal evolution.

In order to extract the semi-analytical relations, we simulated 600 showers with CORSIKA and the CoREAS plugin for the radio emission. Half of them were initiated by proton primaries, while the other half was simulated using iron primaries. One third of the showers had a primary energy of $10^{17} \eV$, another third $10^{18} \eV$ and the final third $10^{19} \eV$. For now, we only use vertical showers (i.e. zenith, $\theta = 0^{\circ}$) and the standard US atmosphere as parametrised by Keilhauer. 

\section{Air shower universality}

The template synthesis method is based on the concept of air shower universality \cite{Lafebre2009}. It is best appreciated when considering the Gaisser-Hillas function \cite{GH77} for the longitudinal evolution of an air shower,

\begin{equation}
	N(X) = N_{\text{max}} \cdot 
	\left( 
		\frac{X - X_1}{X_{\text{max}} - X_1}
	\right)^{
				\frac{X_{\text{max}} - X_1}{\Lambda}
			}
	\cdot \exp
	\left(
		\frac{X_{\text{max}} - X}{\Lambda}
	\right)
	\; .
\end{equation}
Here $X$ represents atmospheric depth in [$\text{g}/\text{cm}^2$] and $N$ the number of particles. The depth of first interaction is denoted as $X_1$ and the subscript ``max'' indicates the values at shower maximum. Lastly, the $\Lambda$ parameter encodes what we will refer to as the shower-to-shower fluctuations. It has been shown that this parameter is universal within 5-10\%, in particular for the electromagnetic part of the shower \cite{Lipari2009}.

With template synthesis, we normalise the radio emission from the air showers with respect to the depth of shower maximum \Xmax, the particle number of shower maximum $N_{\text{max}}$ and the depth of first interaction $X_1$. As the shower-to-shower fluctuations (those encoded by the $\Lambda$ parameter in the longitudinal profile) are not taken into account, we expect the accuracy of template synthesis to be of the same order as these fluctuations.

\newpage
\section{Atmospheric slicing}

\noindent
\begin{minipage}[t]{0.45\textwidth}
	In order to apply the template synthesis method we divide the atmosphere into slices of constant atmospheric depth, as shown in Figure \ref{fig:slicing}. In \cite{Butler:2019xB} it was found that $5 \gcm$ is an optimal value for this spacing, as it is small enough to approximate the individual slices as point sources, but still large enough to achieve good computational performance.
	
	We consider the radio emission coming from each slice separately. This can easily be done using the \textit{slantdepth} keyword in the CoREAS antenna file, in order to configure the antennas to only accept radiation from a given range in slant depth. 
	
	In each slice the radiation scales to first order with the number of particles $\slice{N}$ in that slice. This can be interpreted as a dependence on the primary energy. However, as shown in \cite{Butler:2019xB}, we also need to correct for shower age effects on the pulse shape. We do this by fitting the frequency-depedent amplitude spectrum of the emission, according to a semi-analytical parametrization. The parameters of this function are dependent on the \Xmax -value of the shower, which for a given slice depth $X$ is a proxy for the age of the shower in that slice.
\end{minipage}
~
\begin{minipage}[t]{0.5\textwidth}
	\vspace{-5mm}
	\centering
	\includegraphics[width=0.9\textwidth]{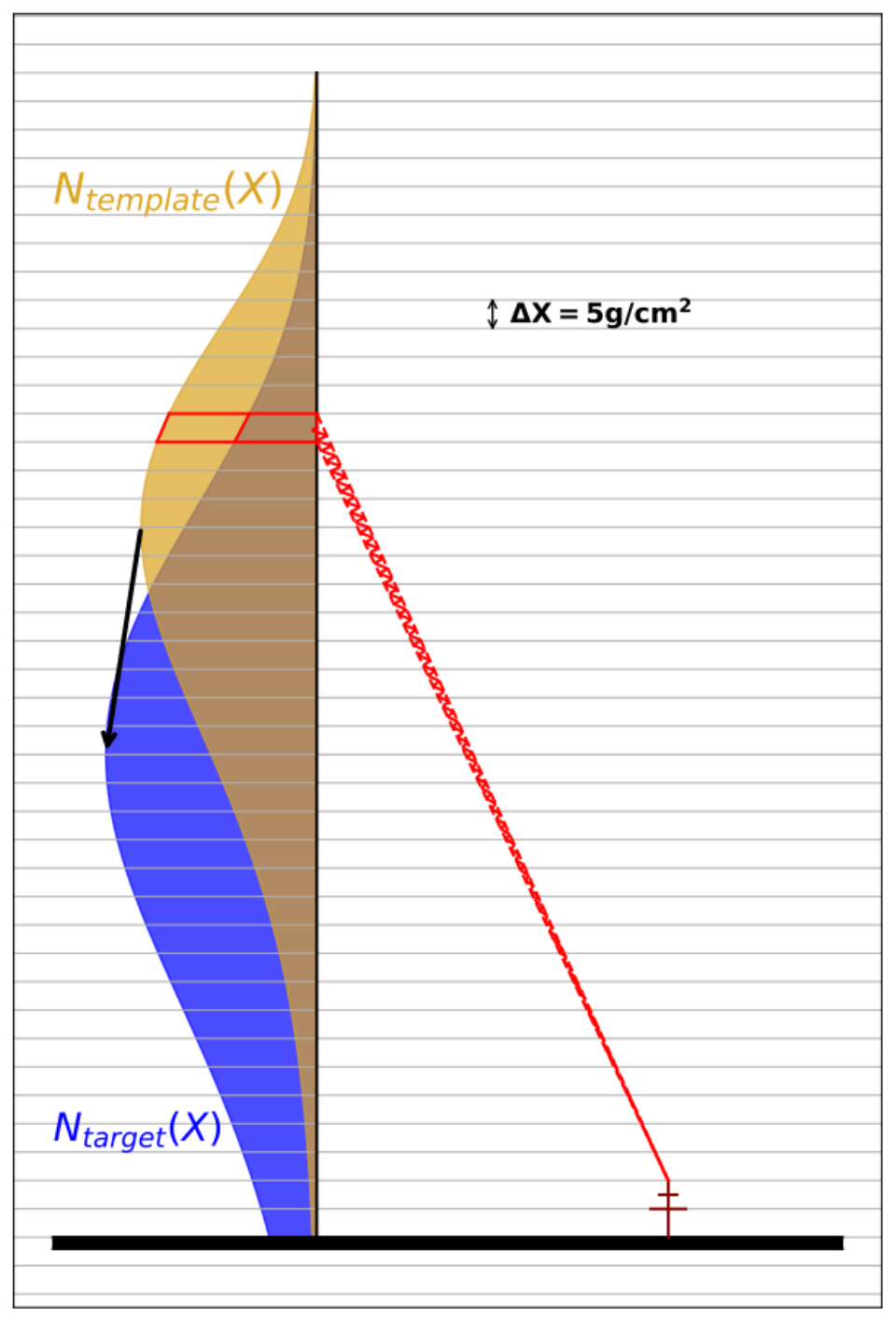}
	\captionof{figure}{The atmospheric slicing performed for template synthesis. The vertical dimension represents the depth in the atmosphere, with the bold horizontal line being ground. The red lines represent the emission from the selected slice. In our setup the x-axis was aligned along the $\vec{v} \times \vec{v} \times \vec{B}$ axis. When scaling the emission from $N_{\text{template}}$ to $N_{\text{target}}$, both the number of particles in the slice as well as the shower age are taken into account.}
	\label{fig:slicing}
\end{minipage}

\section{The template synthesis procedure}

In order to extract the semi-analytical relations, we use 600 vertical showers simulated using CORSIKA v7.6400 with QGSGETII-04 and FLUKA as the interaction models, with parameters as describe above.

The emission was obtained using CoREAS. The antennas were spaced out on the $\vec{v} \times \vec{v} \times \vec{B}$ axis, in order to decouple the geomagnetic and charge excess contributions using their polarization. The antenna positions were chosen to sample the regions in-, outside and on the Cherenkov ring. For each antenna we separate the radiation coming from the different slices. In each antenna, we analyse the emission from each slice as a function of the shower parameters.

Throughout our analysis we used the unscaled version of the discrete Fourier transform, i.e. one needs to scale with the number of samples when going back to the time domain\footnote{This has been the default behaviour in Numpy since at least version 1.10}.

\subsection{Single antenna procedure} \label{subsec:expressions}

In the following we outline the procedure for the case of a single antenna. This can be applied to every antenna independently. At a later stage, we intend to combine template synthesis with an interpolation code to synthesise the emission at arbitrary positions.

The first step is to extract the frequency-dependent amplitude spectra, by applying a real-valued (discrete) Fourier transform to each slice. The resulting spectrum is then parametrised as

\begin{equation} \label{eq:spectrum}
	\tilde{A} (f, \slice{X}) =
	\frac{a}{\slice{N}} \cdot \exp (b \cdot (f - f_0) + c \cdot (f - f_0)^2) + d \; ,
\end{equation}
where $f$ denotes frequency in MHz, $\slice{X}$ is the atmospheric depth at the bottom of the slice and $\slice{N}$ is the number of particles at $\slice{X}$. Because the signal vanishes at zero frequency, we introduce a shift $f_0$ in the frequency domain in order to have better handle on $a$.
The $a$, $b$ and $c$ we call the ``spectral parameters'' and are the free parameters when fitting the spectra. Lastly $d$ represents the noise floor, either due to thinning or to ```particle noise'', and is also parametrised, as

\begin{equation}
	\sqrt{d} = \max 
	\left[
		10^{-9} \cdot 
		\left(
			\frac{\slice{X}}{400 \gcm} - 1.5
		\right) \cdot \exp
		\left( 
			1 - \frac{r_{\text{ant}}}{400 \, \text{m}}
		\right), 0
	\right]  \; ,
\end{equation}
with $r_{\text{ant}}$ being the distance from the antenna to the shower axis.

An important choice one has to make is over which frequency interval to fit this function. As mentioned before, the signal goes rapidly to zero when the frequency goes to zero. From our tests, we saw that the lower frequency part introduces difficulties when fitting, while not contributing much to the physical signal. Therefore, in the following, we will always limit the amplitude spectra to the range $[20, 500] \MHz$, which covers the typical frequency bands of interest. We also set $f_0 = 50 \MHz$ in Equation \eqref{eq:spectrum}. 

\begin{figure}
	\centering
	\includegraphics[width=0.98\textwidth]{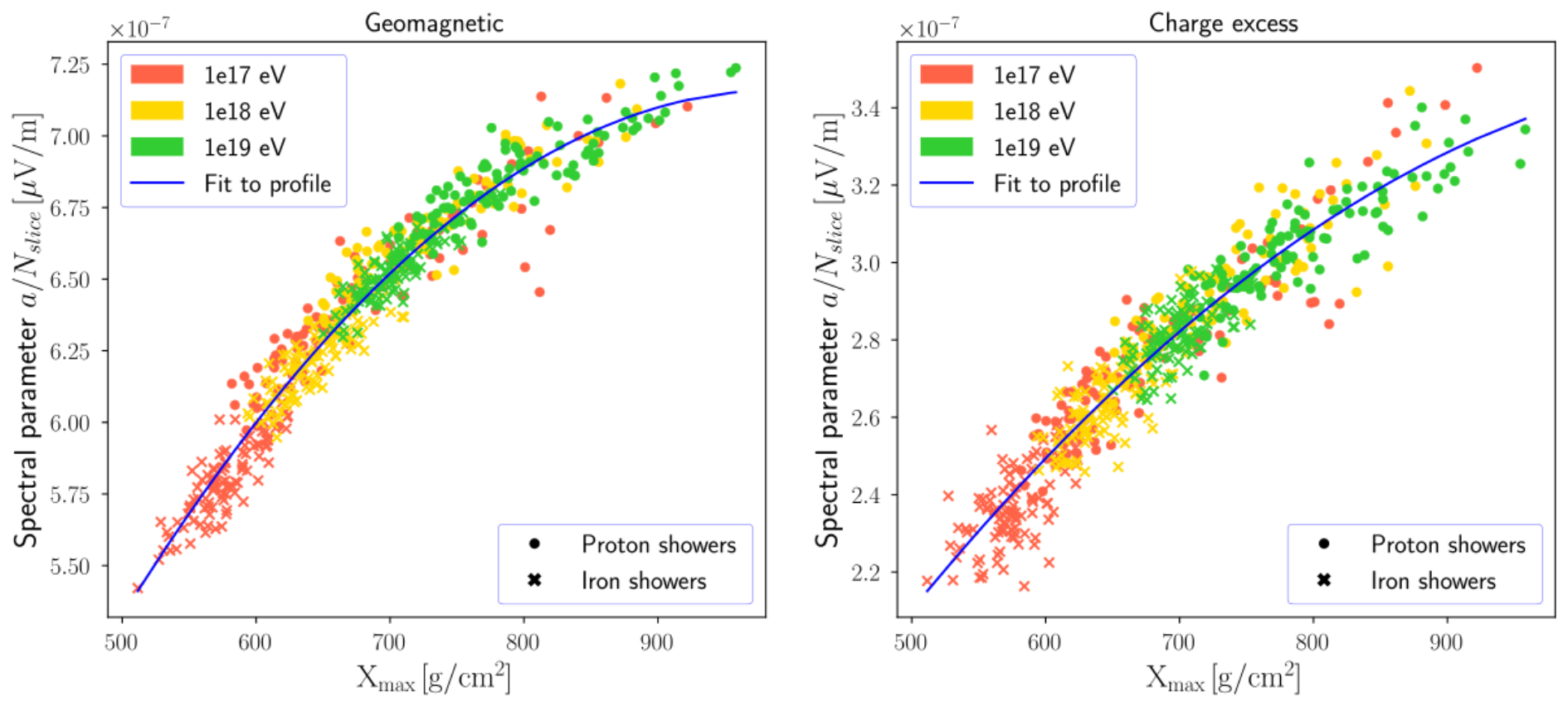}
	\caption{The first spectral parameter from Equation \eqref{eq:spectrum}, $a$, scales quadratically with the \Xmax\ of the shower in each slice, after normalising out the particle number $\slice{N}$. Here we show the case of the slice at $650 \gcm$. This behaviour is independent of the primary energy. These plots show the  $\vec{v} \times \vec{B}$ (left) and  $\vec{v} \times \vec{v} \times \vec{B}$ (right) polarizations for an antenna sitting on the $\vec{v} \times \vec{v} \times \vec{B}$ axis at $110 \;$m from the shower axis. For a vertical shower these correspond to the geomagnetic and charge-excess contributions respectively. Dots indicate proton primaries compared to the crosses which are for the iron induced showers, while the colours reflect the primary energy. The parabolic fit to the distribution is also shown.}
	\label{fig:spectralA}
\end{figure}

\begin{figure}
	\centering
	\includegraphics[width=0.98\textwidth]{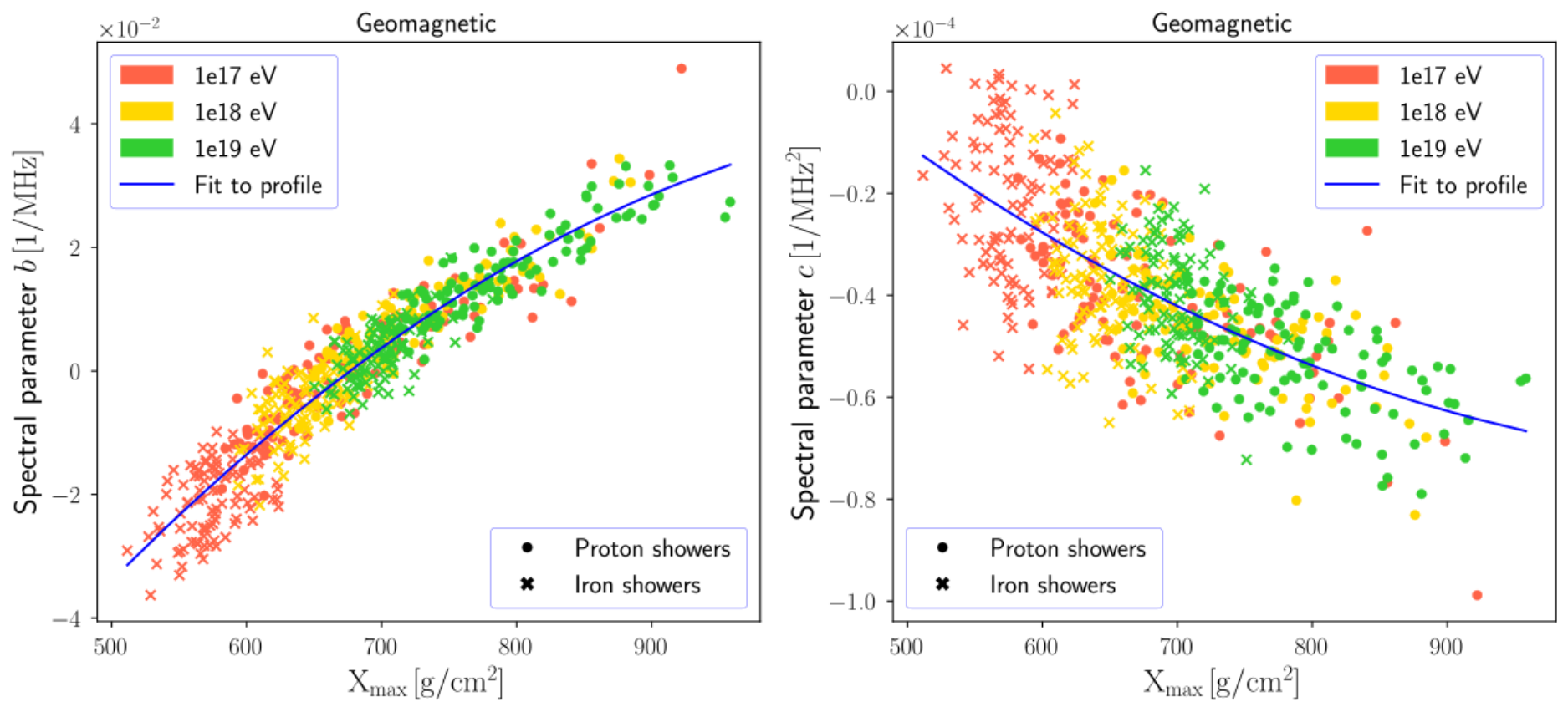}
	
	\caption{The behaviour of the two other spectral parameters from Equation \eqref{eq:spectrum}, for the same combination of antenna and slice as Figure \ref{fig:spectralA}. Although note that here we only show the  $\vec{v} \times \vec{B}$ polarization, corresponding to the geomagnetic component.	While the parabolic behaviour is still present, there is a much higher spread. This can be caused by shower-to-shower fluctuations, but is also influenced by how strongly the spectrum depends on the parameter (as it impacts the fit quality). The spread does not impact the reconstruction quality significantly, as we observe in Section \ref{seq:reconstruction}.}
	\label{fig:spectralBC}
\end{figure}

When we plot the spectral parameter $a$ against the \Xmax\ of the shower in Figure \ref{fig:spectralA}, we observe a quadratic dependence in each polarization separately, independent of primary energy. Similarly for the other spectral parameters in Figure \ref{fig:spectralBC}, although we note that especially for $c$ the spread increases a lot. We believe this is related to the weaker dependence of the signal to this parameter, also indicated by the small values. As we show later, the spread does not impact the reconstruction significantly.

After fitting these distributions we are left with six parabolic curves for every slice, describing the dependence of the spectral parameters on \Xmax\ for both polarizations in the antenna.

\subsection{Synthesising the radio emission from an EAS}

To synthesise the emission, we start from a single microscopic simulation, which we will refer to as the ``origin shower''. After applying a real-valued FFT to the signal in every antenna, we extract the frequency-dependent amplitude spectrum, as well as the phase spectrum. We normalise the amplitude spectrum using the semi-analytical dependencies extracted in Section \ref{subsec:expressions}. This gives us the template amplitude and phase spectra, the latter simply being the one from the origin shower,

\begin{align*}
	A_{\text{template}} (r_{\text{ant}}, f, \slice{X}) 
	&= A_{\text{origin}}(r_{\text{ant}}, f, \slice{X}) \cdot 
	\left[ 
		\slice{N}^{\text{origin}} \cdot \slice{\tilde{A}} (f, X_{\text{max}}, \slice{X})
	\right]^{-1} \\
	\phi_{\text{template}} (r_{\text{ant}}, f, \slice{X})
	&= \phi_{\text{origin}} (r_{\text{ant}}, f, \slice{X})
\end{align*}
These expressions depend on the antenna position $r_{\text{ant}}$. The parametrised spectrum $\tilde{A}$ acquires an \Xmax\ dependence through the spectral parameters. We have also added a superscript to the particle number, to explicitly indicate this is the number of particles in the origin shower.

Given the desired longitudinal profile of the target shower, its signal can now be synthesised. Using the spectral parameters again, we can scale the template according to the \Xmax\ of the target.

\begin{align*}
	A_{\text{target}} (r_{\text{ant}}, f, \slice{X}) 
	&= A_{\text{template}}(r_{\text{ant}}, f, \slice{X}) \cdot \slice{N}^{\text{target}} \cdot \slice{\tilde{A}} (f, X_{\text{max}}, \slice{X}) \\
	\phi_{\text{target}} (r_{\text{ant}}, f, \slice{X})
	&= \phi_{\text{template}} (r_{\text{ant}}, f, \slice{X})
\end{align*}
As such, we can calculate the frequency amplitude and phase spectra for each slice. Summing the slices in the time domain then yields the synthesised pulse in the antenna.

\section{Scoring metrics for the reconstruction} \label{seq:reconstruction}

\begin{figure}
	\centering
	\begin{subfigure}{\textwidth}
		\includegraphics[width=0.95\textwidth]{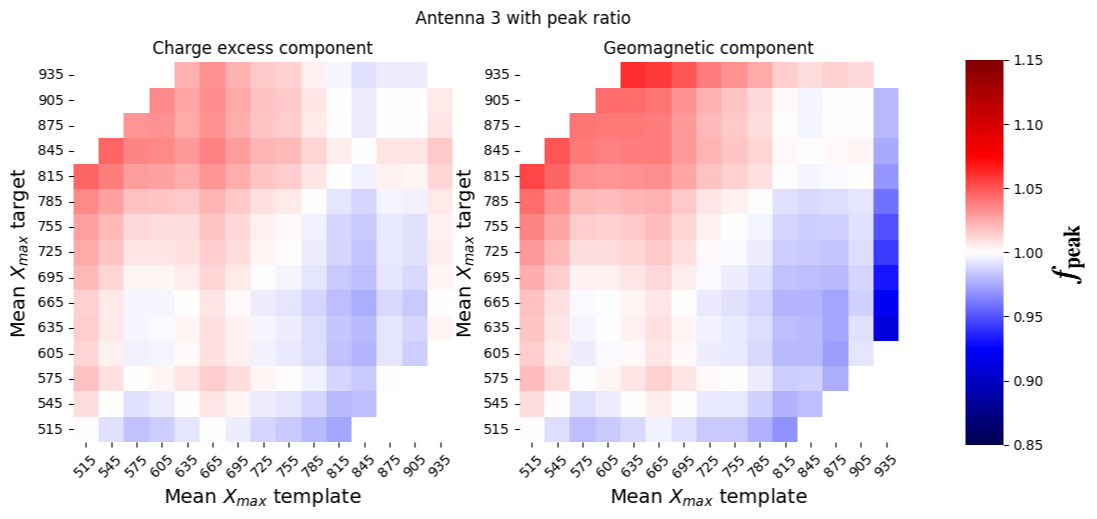}
		\caption{Scores using the $f_{\text{peak}}$ metric, as in Eqn. \eqref{eq:peak_ratio}, which amounts to taking the ratio of the peaks from synthesised and target signals.}
		\label{fig:peak_ratio}
	\end{subfigure} \\
	\begin{subfigure}{\textwidth}
		\includegraphics[width=0.95\textwidth]{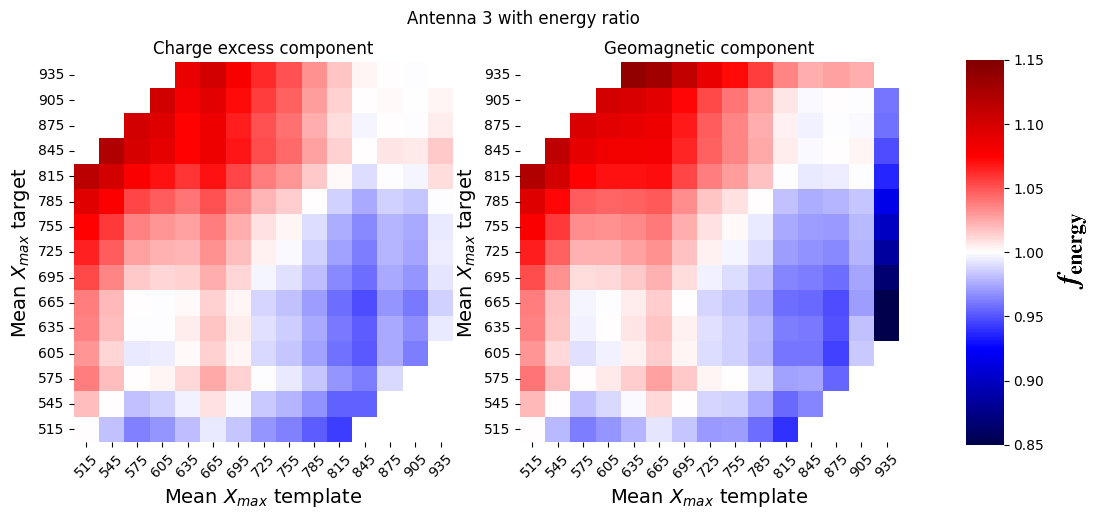}
		\caption{Scores using the $f_{\text{energy}}$ metric, as in Eqn. \eqref{eq:energy_ratio}, which takes the ratio of the fluences in synthesised and target signals.}
		\label{fig:energy_ratio}
	\end{subfigure}
	\caption{We use each shower in our simulation set as a template to synthesise the signal from several other target showers in the set. The synthesised signal is compared with the signal from CoREAS and scored using different metrics, for each polarization separately. The scores are binned according to the \Xmax\ of template and target. 
	We only synthesised showers with $ \Delta X_{\text{max}} \leq 300 \gcm$, as we don't expect using this technique on larger differences. Here we show the scores for  antenna 3, which sits at $110 \;$m from the shower core on the $\vec{v} \times \vec{v} \times \vec{B}$ axis.}
\end{figure}

To evaluate how well our method synthesises the radio signals, we implemented several scoring metrics to compare the synthesised traces $s_{\text{synth}}(t)$ with those coming from CoREAS, $s_{\text{reas}}(t)$. The first metric takes the ratio of the maxima,

\begin{equation} \label{eq:peak_ratio}
	f_{\text{peak}}(s_{\text{synth}}, s_{\text{reas}}) = \frac{\max(s_{\text{synth}}(t))}{\max(s_{\text{reas}}(t))} \; .
\end{equation}
To gauge the performance over our simulation set, we took each shower as a template to synthesise the emission from several others in the set. We scored the resulting signals using Equation \eqref{eq:peak_ratio} and binned the results depending on the \Xmax\ of template and target, in order to get an average performance which is not dominated by single shower fluctuations. The result is shown in Figure \ref{fig:peak_ratio}.

Another way to compare the traces, would be to look at the energy fluence contained in each of them. With energy fluence defined as

\begin{equation}
	E(s) \propto \Delta t \sum_{n} s(n \cdot \Delta t)^2 \; ,
\end{equation}
where $\Delta t$ is the time step length in the trace, we can take the ratio as a metric,

\begin{equation} \label{eq:energy_ratio}
	f_{\text{energy}}(s_{\text{synth}}, s_{\text{reas}}) = \frac{E(s_{\text{synth}})}{E(s_{\text{reas}})} \; .
\end{equation}
The scores using this metric are visualised in Figure \ref{fig:energy_ratio}.

It is clear from these results that the quality of the synthesis depends mostly on the $\Delta X_{\text{max}} = X_{\text{max}}^{\text{target}} - X_{\text{max}}^{\text{template}}$. This is unsurprising, as the larger this difference, the further we extrapolate the emission using the spectral parameters. Further studies will have to show how far we can take this extrapolation, but from this first approach we can already conclude that we should keep $ \Delta X_{\text{max}} \leq 300 \gcm$ to synthesise with amplitude deviations better than 10\%.

Nonetheless, for both metrics we find an asymmetric distribution. 
We will investigate this behaviour further, once we have generalised the method to arbitrary geometries. For now we note that overall we synthesise the signals better than 10-15\% for both metrics, which indeed corresponds to the expected level of the shower-to-shower fluctuations. 

\section{Conclusion}

We presented the template synthesis method, a novel way to simulate the emission from extensive air showers. It uses a single microscopic simulation, sliced into chunks of fixed atmospheric depth, to create a template. Using semi-analytical expressions, which describe the dependence of the frequency-dependent amplitude spectrum in every slice on the \Xmax\ and thus age of the shower, we can rescale the emission from the template to synthesise any other shower. In this contribution, we applied this method to vertical air showers.

The semi-analytical expressions were extracted from a simulation library of 600 CORSIKA showers. For each of these, we parametrised the normalised frequency-dependent amplitude spectra in every slice, for every antenna separately. The spectral parameters of this function exhibit a quadratic dependence on \Xmax , a fact we could use to find the average amplitude spectrum for a given slice and antenna.

Using two metrics, we scored the synthesised signals. We found the synthesis quality to be most sensitive to the difference in \Xmax\ between template and target. Furthermore, the method shows an asymmetry between going from low to high \Xmax\ and vice versa. While this could definitely be improved, we want to first explore other geometries. Lastly we note for both metrics we achieved a reconstruction quality better than 10-15\%, which corresponds to the shower-to-shower fluctuations we expect to be limited by. We conclude this method works well for vertical air showers.

\bibliographystyle{JHEP}
\bibliography{bibliography}

\end{document}